\documentclass[aip,jap,reprint,showkeys,twocolumn,nofootinbib]{revtex4-1}
\bibpunct{[}{]}{,}{n}{}{}
\usepackage{natbib}
\usepackage{amssymb}
\usepackage{color}
\usepackage{amsfonts}
\usepackage{textcomp}
\newcommand{\ped}[1]{\ensuremath{_{\rm #1}}}
\newcommand{\apex}[1]{\ensuremath{^{\rm #1}}}
\newcommand{\parallelslant}{\mathbin{\!/\mkern-5mu/\!}}
\definecolor{blue}{rgb}{0,0,0}
\usepackage{float}
\usepackage[caption = false]{subfig}
\usepackage{graphicx}

\begin{document}

\title{Two-dimensional hole transport in ion-gated diamond surfaces{\color{blue}: A brief review}}

\author{Erik Piatti}
\author{Davide Romanin}
\author{Dario Daghero}
\author{Renato S. Gonnelli}
\affiliation{\mbox{Department of Applied Science and Technology, Politecnico di Torino, I-10129 Torino, Italy}}

\begin{abstract}
Electrically-conducting diamond is a promising candidate for next-generation electronic, thermal and electrochemical applications. One of the major obstacles towards its exploitation is the strong degradation that some of its key physical properties -- such as the carrier mobility and the superconducting transition temperature -- undergo upon the introduction of disorder. This makes the two-dimensional hole gas induced at its surface by electric field-effect doping particularly interesting from both a fundamental and an applied perspective, since it strongly reduces the amount of extrinsic disorder with respect to the standard boron substitution. {\color{blue}In this short review, we summarize} the main results achieved so far in controlling the electric transport properties of different field-effect doped diamond surfaces via the ionic gating technique. We analyze how ionic gating can tune their conductivity, carrier density and mobility, and drive the different surfaces across the insulator-to-metal transition. We review their strongly orientation-dependent magnetotransport properties, with a particular focus on the gate-tunable spin-orbit coupling shown by the (100) surface. Finally, we discuss the possibility of field-induced superconductivity in the (110) {\color{blue}and (111) surfaces} as predicted by density functional theory calculations.
\end{abstract}


\maketitle

\section*{Introduction}\label{sec:introduction} 

Intrinsic diamond is a wide-bandgap insulator that has attracted significant interest from both a fundamental and an applied perspective, owing to its ultrahigh thermal conductivity, large intrinsic charge-carrier mobility, high breakdown electric field, excellent electrochemical stability, and biocompatibility \cite{PanBook}. One of the key challenges to exploit its potential lies in how to induce a large electrical conductivity without an excessive degradation of its excellent intrinsic transport properties. Hole-type conductivity can be induced in bulk diamond by substituting boron (B) to carbon in the lattice structure, inducing an insulator-to-metal transition and superconductivity (SC) at low temperature \cite{EkimovNature2004, BustarretPRL2004, YokoyaNature2005}. However, the necessary dopant concentrations are very large ($> 10^{20}\,\mathrm{cm^{-3}}$) and introduce significant lattice disorder, thus being detrimental both for the normal-state and SC transport properties \cite{TakanoAPL2004, IshizakaPRL2007, BustarretPSSA2008, OkazakiAPL2015}. Disorder and limited B solubility have also hampered the exploitation of the ultrahigh Debye temperature of diamond to achieve the predicted high-temperature SC \cite{BoeriPRL2004, LeePRL2004, BoeriJPCS2006, GiustinoPRL2007}.

The surface of diamond can exhibit hole-type conductivity when it is functionalized with hydrogen and exposed to electron-accepting molecules (e.g., air moisture) for at least few tens of minutes \cite{LandstrassAPL1989, MaierPRL2000, StrobelNature2004} due to an electron-transfer process between the valence band of diamond and the available energy levels of the adsorbed molecules \cite{MaierPRL2000, StrobelNature2004}. The electric transport properties of this surface conducting layer have been investigated extensively in the literature \cite{LandstrassAPL1989, MaierPRL2000, StrobelNature2004, KawaradaSSR1996, NebelDRM2002, NebelDRM2004, EdmondsPRB2010, EdmondsNanoLett2015} and exploited to fabricate high-performance transistors and bionsensors \cite{HiramaAPL2008, DankerlAFM2009, HiramaAPE2010, KawaradaPSSA2011}. The surface hole densities induced by this charge-transfer process are modest ($\lesssim 10^{13}\,\mathrm{cm^{-2}}$), but already at the limits attainable by means of standard transistor configurations. In recent years, a much wider range of tunability (up to $\sim 10^{15}\,\mathrm{cm^{-2}}$) was achieved by means of the ionic gating technique \cite{FujimotoPCCP2013, UenoJPSJ2014}, where the ultrahigh capacitance of an electrolyte/electrode interface is exploited to modulate the charge carrier density and strongly modify the physical properties of the gated material \cite{YeScience2012, DagheroPRL2012, YeNatMater2010, ShiSciRep2015, YuNatNano2015, LiNature2016, WangNature2016, XiPRL2016, ShiogaiNatPhys2016, PiattiJSNM2016, ZengNanoLett2018, DengNature2018, WangNatNano2018, PiattiPRM2019}. While the first reports of electrolyte-gated diamond-based transistors were focused on chemical and biological sensing \cite{KawaradaPSSA2001, NebelJAP2006, DankerlPRL2011, DankerlAPL2012}, several later investigations appeared in the literature where the electric transport properties of the gate-induced two-dimensional hole system were front and center \cite{YamaguchiJPSJ2013, TakahidePRB2014, AkhgarNanoLett2016, TakahidePRB2016, AkhgarPRB2019, PiattiEPJST2019, NakamuraPRB2013, SanoPRB2017, RomaninApSuSc2019}.

In this short review, we present the main results obtained so far concerning the electric transport properties of ion-gated diamond, both from the theoretical and experimental point of view, with a specific focus on the behavior shown by the different diamond surfaces at low temperature. In Section \ref{sec:bandstructure}, we describe the electronic bandstructure of (111) and (100) hydrogenated diamond, which are the only two single-crystal surfaces experimentally investigated so far. In Section \ref{sec:ionic_gate}, we describe how the ionic gate technique applied to the (111), (100) and nanocrystalline diamond (NCD) surfaces allows tuning their high-temperature electric transport properties. In Section \ref{sec:IMT}, we present an analysis of the gate-induced insulator-to-metal transition (or lack thereof) in the same surfaces. In Section \ref{sec:low_T}, we focus on the strongly orientation-dependent magnetotransport properties of the (111) and (100) surfaces at low temperature. In Section \ref{sec:supercond}, we examine the theoretical predictions for gate-induced SC in the (110) {\color{blue}and (111) surfaces}. Finally, in Section \ref{sec:summary} we summarize the main results and provide a possible outlook for future investigations.

\section{Electronic bandstructure}\label{sec:bandstructure}

When a bulk crystal is cleaved along one of its structural crystallographic planes, the exposed surface becomes very reactive to the presence of both neighboring sites and adsorbates. As a consequence, the cleaved surface may either undergo a reconstruction, or the dangling bonds may be saturated by bonding other atomic species. Intrinsic diamond is often grown via chemical vapor deposition (CVD) in a hydrogen-rich atmosphere, so that hydrogen termination is the most natural choice to passivate the exposed surfaces. Here we review the electronic bandstructure of the hydrogen-terminated, (111)- and (100)-oriented diamond surfaces as calculated by ab-initio density functional theory (DFT) using various methods \cite{SquePRB2006, RiveroC2016, KernSurfSc1996, KernSurfSc1996_b, PamukPRB2019}.

\begin{figure}
\begin{center}
\includegraphics[keepaspectratio, width=1.0\columnwidth]{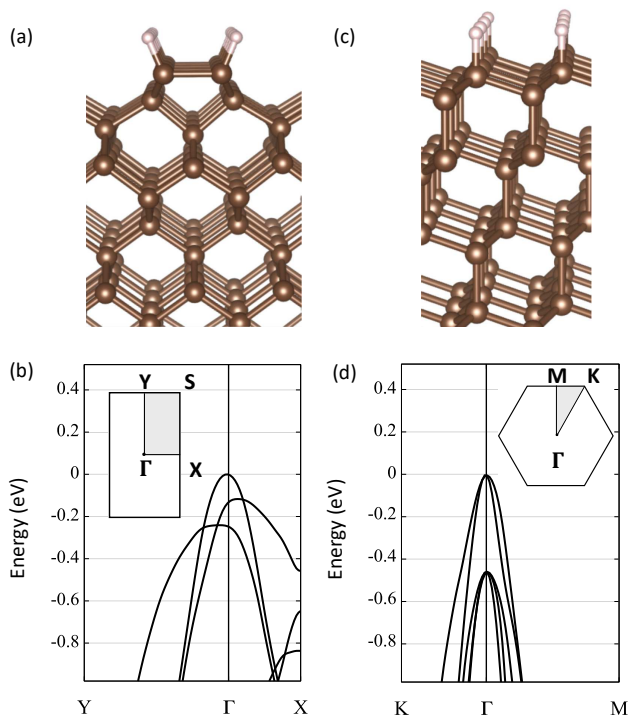}
\end{center}
\caption {
(a) Ball-and-stick model of the hydrogen-terminated, reconstructed (100) $2\times1$  diamond surface, and 
(b) the relevant electronic bandstructure.
(c,d) Same as (a,b) for the hydrogen-terminated, unreconstructed (111) $1\times1$  diamond surface. Large brown spheres and small pink spheres are carbon and hydrogen atoms respectively. The zero of the energy is set to the valence-band maximum. The insets to (b) and (d) represent the first Brillouin zone. Grey shaded areas are the irreducible Brillouin zones. Data in (b,d) are adapted from Ref.\onlinecite{SquePRB2006}.
} \label{figure:bands}
\end{figure}

The as-cleaved (100) diamond surface features two dangling bonds per surface carbon atom, which cannot be entirely saturated by hydrogen due to steric repulsion between adsorbates \cite{FrauenheimPRB1993, JingSurfSc1994}. As a consequence, the hydrogenated $1\times1$ surface is unstable and reconstructs into the $2\times1$ supercell shown in Fig.\ref{figure:bands}a. This reconstructed surface exhibits dimer rows between neighboring carbon atoms and covalent bonds between hydrogen and carbon. The surface point group is $C_{2v}$ where [100] is a two-fold rotation axis, while there are two mirror planes orthogonal to the [011] and [01$\bar{1}$] axes. The corresponding bandstructure is shown in Fig.\ref{figure:bands}b. The uppermost three valence bands arise from bulk states, since the occupied surface states lie $\approx 2\,\mathrm{eV}$ below the top of the {\color{blue}bulk} valence band, while the empty surface states lie $\approx 3\,\mathrm{eV}$ above the highest occupied {\color{blue} band} \cite{KernSurfSc1996}. In the absence of external doping, the hydrogen-terminated $2\times1$ (100) surface is thus insulating with a DFT-calculated band gap of {\color{blue} $\approx 3.0\,\mathrm{eV}$ \cite{SquePRB2006}}.

The as-cleaved (111) diamond surface features only one dangling bond per surface carbon atom, making the $1\times1$ surface cell stable upon hydrogenation of the missing bonds. In the absence of hydrogenation, the clean surface reconstructs in the $2\times1$ supercell instead, forming carbon dimers \cite{PandeyPRL1981, PamukPRB2019}. The hydrogenated (111) surface {\color{blue}is shown in Fig.\ref{figure:bands}c}. The corresponding point group is $C_{3v}$ where [111] is a three-fold rotation axis and there are three reflection planes (orthogonal to the [2$\bar{1}\bar{1}$], [$\bar{1}\bar{1}$2] and [$\bar{1}$2$\bar{1}$] axes). Similarly to the $2\times1$ (100) surface, the hydrogenated $1\times1$ (111) diamond surface is insulating with a direct band gap (calculated by DFT) of {\color{blue}$\approx 2.5\,\mathrm{eV}$ \cite{SquePRB2006}}, its uppermost valence bands originate from bulk states, and its occupied surface states are $\approx 4\,\mathrm{eV}$ below the top of the {\color{blue}bulk} valence band.

Hydrogenation of the diamond surfaces stabilizes them against reconstruction and makes them insulating. However, it also lowers their electron affinities and ionization energies due to the polarity of adatom-substrate bonds, which can be treated as heteropolar molecules \cite{RisteinAPA2006, SquePRB2006, RiveroC2016}. This makes these surfaces extremely susceptible to charge transfers upon adsorption of electron-accepting species: Hydrogen redox extracts electrons from the valence bands of diamond, leaving behind surface-bound holonic carriers which are mobile in the in-plane directions and lead to a finite surface electrical conductivity \cite{LandstrassAPL1989, MaierPRL2000, StrobelNature2004}. For simplicity, in the following we shall refer to the reconstructed $2\times1$ (100) hydrogenated diamond surface and to the unreconstructed $1\times1$ (111) hydrogenated diamond surface as the (100) and (111) surfaces respectively.

\section{Ionic-gate operation}\label{sec:ionic_gate}

\begin{figure*}
\begin{center}
\includegraphics[keepaspectratio, width=0.75\textwidth]{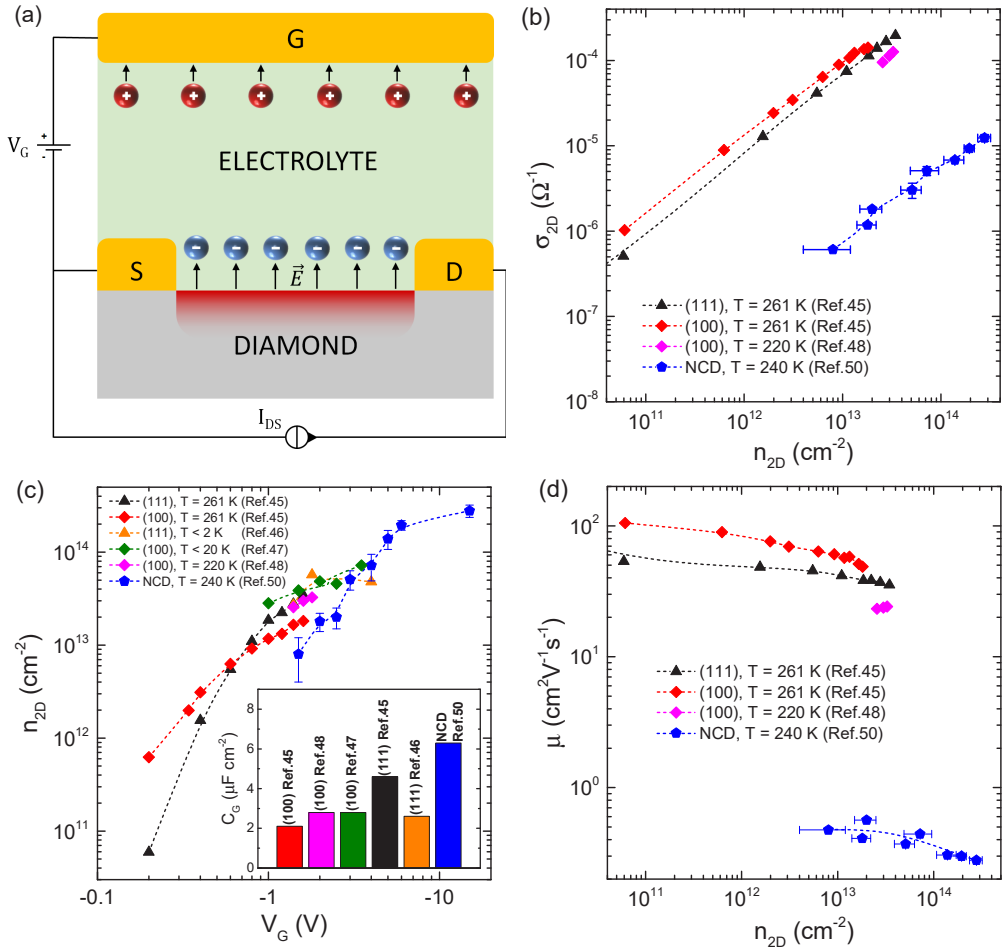}
\end{center}
\caption {
(a) Sketch of a diamond-based EDL transistor biased with a negative $V_G$. Electric field $\vec{E}$ and source (S), drain (D) and gate (G) electrodes are indicated. The gate-induced 2DHG is represented by the shaded red area.
(b) In-plane sheet conductance in the accumulation layer $\sigma_{2D}$ as a function of the induced hole density $n_{2D}$, in different gated diamond surfaces at high $T$. 
(c) Induced hole density $n_{2D}$ as a function of $V_G$ in different gated diamond surface. The inset shows the EDL capacitance $C_G$ extrapolated in different papers from the range where $n_{2D}$ scales linearly with $V_G$.
(d) High-$T$ mobility $\mu$ as a function of $n_{2D}$ in different gated diamond surfaces.
Surface type and $T$ at which $V_G$ was applied are reported in the legend in all panels. Dashed lines are all guides to the eye.
Data are adapted from Refs.\onlinecite{YamaguchiJPSJ2013, TakahidePRB2014, AkhgarNanoLett2016, TakahidePRB2016, PiattiEPJST2019}.
} \label{figure:EDLT}
\end{figure*}

It is well known that the application of an electric field perpendicular to the surface of a semiconductor can result in the formation of a two-dimensional gas of charge carriers with a tunable density. The ionic gating technique can be employed to induce the electric field effect in the electric double-layer (EDL) transistor configuration, which we sketch in Fig.\ref{figure:EDLT}a for an ion-gated diamond surface. Here, the diamond surface is separated from a metal gate electrode (G) by an electrolyte, which can be an acqueous solution \cite{DankerlPRL2011}, a solid polymer electrolyte \cite{DankerlAPL2012}, an ionic liquid \cite{YamaguchiJPSJ2013, TakahidePRB2014, AkhgarNanoLett2016, TakahidePRB2016, AkhgarPRB2019} or an ion-gel \cite{PiattiEPJST2019}. When a negative voltage $V_G$ is applied to the gate electrode, negative ions accumulate at the diamond surface and induce the formation of the two-dimensional hole gas (2DHG), thus building up the EDL. Typically, $V_G$ is applied immediately above the freezing point of the chosen electrolyte to minimize the chance of unwanted electrochemical reactions. The conductivity of the 2DHG can be monitored by applying a small current $I_{DS}$ between the source (S) and drain (D) electrodes and simultaneously measuring the longitudinal voltage drop. 

In Fig.\ref{figure:EDLT}b we show the high-temperature conductivity per unit surface, $\sigma_{2D}$, as a function of the carrier density $n_{2D}$ for different ion-gated diamond surfaces. In the (111) and (100) hydrogenated single-crystal surfaces \cite{YamaguchiJPSJ2013, TakahidePRB2016}, $\sigma_{2D}$ can be controlled over nearly three orders of magnitude upon the induction of moderate carrier densities $n_{2D}\lesssim 4\cdot 10^{13}\,\mathrm{cm^{-2}}$. The $\sigma_{2D}$ of the accumulation layer induced at the much rougher surface of B-doped NCD films is less tunable (slighly less than two orders of magnitude) and requires much larger carrier densities ($10^{13}-10^{14}\,\mathrm{cm^{-2}}$) to be modified \cite{PiattiEPJST2019}. Additionally, the exposure of the hydrogenated surfaces to the electrolyte always results in an initial reduction of $\sigma_{2D}$ by at least one order of magnitude, as the adsorbates responsible for charge-transfer doping dissolve in the electrolyte \cite{YamaguchiJPSJ2013}; whereas the conductivity of the B-doped films is insensitive to the deposition of the electrolyte \cite{PiattiEPJST2019}.

The behavior of the induced carrier density $n_{2D}$ as a function of the applied $V_G$ is also dependent on the diamond surface (see Fig.\ref{figure:EDLT}c). In general, $n_{2D}$ becomes larger for larger negative values of $V_G$, with at least one voltage range -- usually at low $V_G$ -- where the increase is linear and the diamond/electrolyte interface capacitance $C_G$ is nearly constant \cite{YamaguchiJPSJ2013, AkhgarNanoLett2016, PiattiEPJST2019}. The exact values and trends are dependent on which electrolyte is used and at which temperature $T$ one measures $n_{2D}$, as is the maximum negative $V_G$ which can be applied before the onset of irreversible electrochemical reactions. The inset to Fig.\ref{figure:EDLT}c summarizes the values of $C_G$ determined by fitting the $n_{2D}$ vs. $V_G$ curves in the linear region. Despite sample-to-sample fluctuations, it can be seen that the (111) hydrogenated surface features a larger capacitance ($C_G \simeq 2.6 - 4.6\,\mathrm{\mu F\,cm^{-2}}$)\cite{YamaguchiJPSJ2013, TakahidePRB2014} than the (100) surface ($C_G \simeq 2.1 - 2.8\,\mathrm{\mu F\,cm^{-2}}$)\cite{YamaguchiJPSJ2013, AkhgarNanoLett2016, TakahidePRB2016}. This difference may be associated to the different density of C\apex{-}-H\apex{+} dipoles between the two surfaces \cite{YamaguchiJPSJ2013, HiramaAPE2010}, or possibly to different values of density of states (DOS) at the Fermi level leading to different values of the quantum capacitance. The even larger value of $C_G \simeq 6.3\,\mathrm{\mu F\,cm^{-2}}$ of the NCD surface is instead likely associated with the finite B content providing a finite carrier density even at $V_G = 0$ \cite{PiattiEPJST2019}, which in turn improves the electrostatic screening with respect to the undoped surfaces.

Knowing both $n_{2D}$ and $\sigma_{2D}$ allows determining the mobility $\mu$ in the various 2DHGs. As we show in Fig.\ref{figure:EDLT}d, the mobilities of the (111) and (100) hydrogenated surfaces are comparable, the values of $\mu$ for the (100) surface being slightly larger for most of the values of $n_{2D}$ (though it could be partially due to sample-to-sample fluctuations) \cite{YamaguchiJPSJ2013, TakahidePRB2016}. On the other hand, the values of $\mu$ for the B-doped NCD surface are two orders of magnitude lower at comparable $n_{2D}$, obviously due to the polycrystalline nature of the samples and their much larger surface roughness ($\sim 30$ nm for NCD vs. $\sim 0.5$ nm for single crystals) \cite{PiattiEPJST2019}. Despite these differences, $\mu$ is always decreasing with increasing $n_{2D}$ for all diamond surfaces. Ref.\onlinecite{YamaguchiJPSJ2013} proposed this behavior to be due to a density-dependent increase in carrier-phonon scattering, in analogy with silicon inversion layers \cite{AndoRMP1982} and ungated hydrogenated diamond \cite{HiramaAPE2010}. In Ref.\onlinecite{PiattiEPJST2019}, instead, the decrease of $\mu$ is associated to the fact that ionic gating itself introduces extrinsic disorder due to the presence of the charged ions in the EDL, leading to suppressed mobilities with respect to standard solid-gating techniques \cite{GallagherNatCommun2015, SaitoACSNano2015, Gonnelli2dMater2017, PiattiApSuSc2017, PiattiNanoLett2018, PiattiApSuSc2018mos2} or even a suppression of metallic charge transport \cite{OvchinnikovNatCommun2016, PiattiAPL2017, LuPNAS2018}. As we will discuss in the next Section, the low-$T$ mobility (where no significant phonon scattering is present) decreases only for sufficiently large values of $n_{2D}$: thus, it is likely that the suppression of $\mu$ with increasing $n_{2D}$ at high T is caused by a combination of both effects.

\section{Insulator-to-metal transition}\label{sec:IMT}

\begin{figure*}
\begin{center}
\includegraphics[keepaspectratio, width=1.0\textwidth]{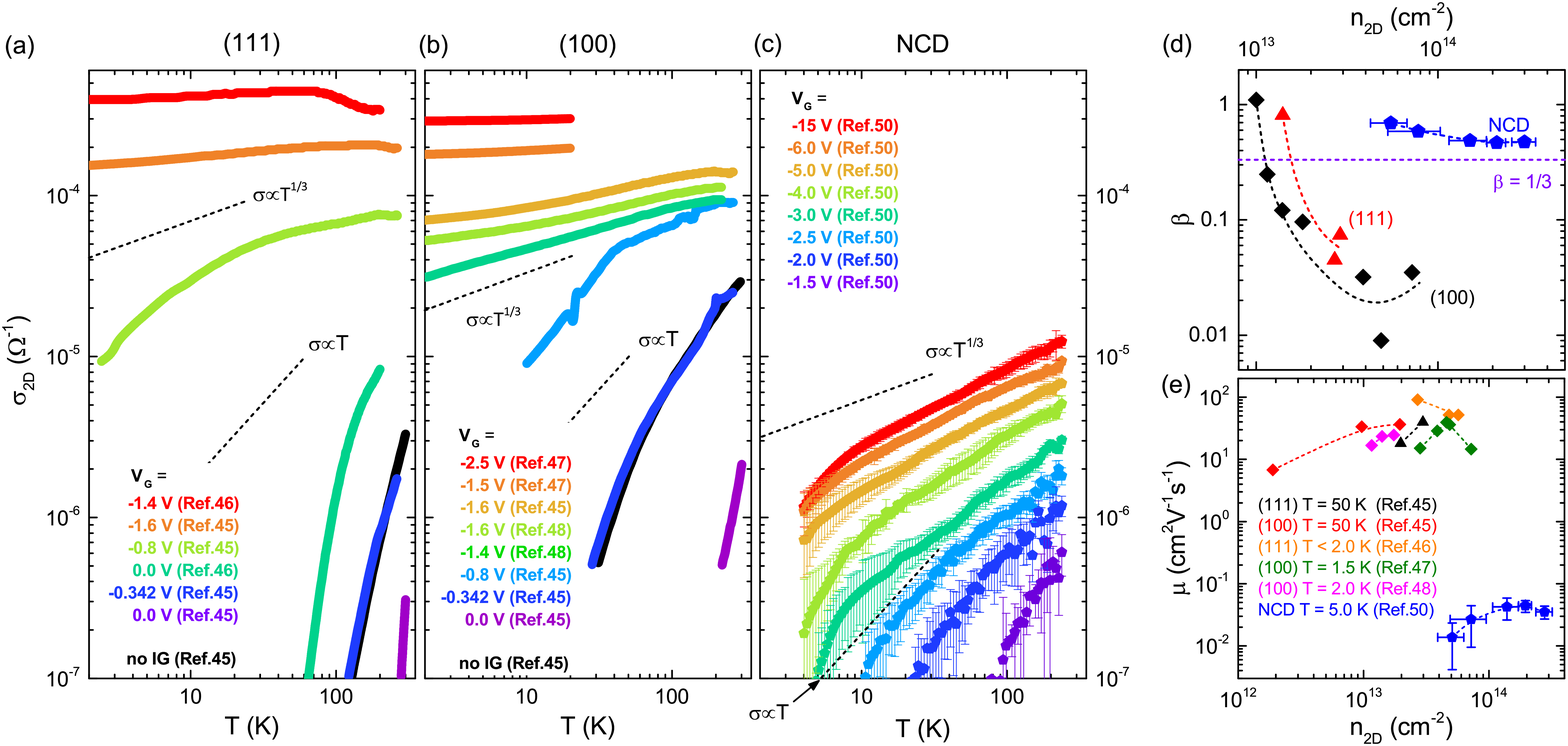}
\end{center}
\caption {
(a) Double logarithmic plot of the in-plane sheet conductance $\sigma_{2D}$ as a function of $T$ in gated (111) diamond surfaces, for different devices and values of $V_G$. Dashed lines mark the scaling at the boundary between the hopping and QC regime ($\sigma \propto T$) and at the QC point of the IMT ($\sigma \propto T^{1/3}$). The label \mbox{``no IG"} marks the curve in absence of the ionic gate.
(b) Same as in (a) for gated (100) diamond surfaces.
(c) Same as in (a) for B-doped NCD surfaces.
(d) Scaling factor $\beta$ of the QC regime for increasing $n_{2D}$ in different gated diamond surfaces. The dashed violet line marks the QC point $\beta = 1/3$. Other dashed lines are guides to the eye.
(e) Low-$T$ mobility $\mu$ as a function of $n_{2D}$ in different gated diamond surfaces. Surface type and $T$ at which $\mu$ was estimated are reported in the legend. Dashed lines are guides to the eye.
Data are adapted from Refs.\onlinecite{YamaguchiJPSJ2013, TakahidePRB2014, AkhgarNanoLett2016, TakahidePRB2016, PiattiEPJST2019}.
} \label{figure:IMT}
\end{figure*}

Tuning the carrier density in the gate-induced 2DHG has a profound impact on its charge conduction mechanism, as evidenced by the $T$-dependence of $\sigma_{2D}$ (see Fig.\ref{figure:IMT}a-c). In the absence of the ionic gate, $\sigma_{2D}$ decreases with decreasing $T$ in both the (111) and (100) hydrogenated surfaces (solid black lines in Fig.\ref{figure:IMT}a and b respectively), mostly due to a strong suppression of the hole mobility \cite{YamaguchiJPSJ2013}. The deposition of the ionic medium suppresses $\sigma_{2D}$ (solid violet lines in Fig.\ref{figure:IMT}a-b) at $V_G = 0$, but the original $\sigma_{2D}$ can be recovered by applying a small $|V_G|<0.4\,\mathrm{V}$ (solid blue lines in Fig.\ref{figure:IMT}a-b). Even though different samples having the same surface termination often exhibit different values of $\sigma_{2D}$ at $V_G = 0$, increasing the hole density by applying negative values of $V_G$ results in an increase of $\sigma_{2D}$ and in a weakening of its $T$-dependence. Indeed, when the applied negative $V_G$ is sufficiently large, $\sigma_{2D}$ becomes nearly $T$-independent for $T\rightarrow0$ in both the (111) and (100) hydrogenated surfaces \cite{YamaguchiJPSJ2013, TakahidePRB2014, AkhgarNanoLett2016}, suggesting a gate-induced insulator-to-metal transition (IMT). Conversely, this low-$T$ saturation is never observed in the NCD B-doped surfaces \cite{PiattiEPJST2019} (see Fig.\ref{figure:IMT}c). No evidence for SC behavior has been observed so far in any ion-gated diamond surfaces down to the lowest measured $T$, most likely the achieved values of $n_{2D}$ are not large enough \cite{YamaguchiJPSJ2013, TakahidePRB2014, AkhgarNanoLett2016, TakahidePRB2016, AkhgarPRB2019, PiattiEPJST2019}.

Further insight in the gate-induced IMT can be gained by considering the scaling of $\sigma_{2D}$ with $T$. When $n_{2D}$ is low enough (the exact values being surface- and sample-dependent), $\sigma_{2D}$ decreases with decreasing $T$ faster than any power law, indicating that charge transport occurs through a hopping mechanism between localized states \cite{HeegerPS2002, MottBook1979, MottBook1990}. In the case of NCD B-doped surfaces, the hopping mechanism has been explicitly identified to be of the 2D Mott variable-range type. Upon increasing the hole density, the low-$T$ behavior of $\sigma_{2D}$ changes into a power law, $\sigma_{2D} \propto T^\beta$, with $\beta \lesssim 1$. This is the hallmark of the quantum critical (QC) regime of the IMT \cite{HeegerPS2002, MottBook1979, MottBook1990}. When $\beta > 1/3$, $\sigma_{2D}$ still vanishes for $T\rightarrow0$, identifying the insulating side of the IMT. Notably, the NCD B-doped surface can never be brought beyond this regime even at the largest values of $n_{2D}$, most likely due to its very low mobility caused by the large surface roughness \cite{PiattiEPJST2019}. On the other hand, sufficiently large values of $n_{2D}$ allow the hydrogenated single-crystal surfaces to cross the $\beta = 1/3$ line (the QC point of the IMT \cite{HeegerPS2002, MottBook1979, MottBook1990}) and reach the metallic regime where a finite $\sigma_{2D}$ exists for $T\rightarrow0$. In this regime, the residual weak $T$-dependence of $\sigma_{2D}$ has been attributed to weak localization in the (111) surface \cite{TakahidePRB2014} and to hole-hole interactions in the (100) surface \cite{AkhgarNanoLett2016}. The $n_{2D}$-dependences of the scaling factor $\beta$ for the various diamond surfaces are summarized in Fig.\ref{figure:IMT}d.

As anticipated in the previous Section, the $n_{2D}$-dependence of the low-$T$ mobility is different from that of the high-$T$ one. Namely, $\mu$ no longer monotonously decreases with increasing $n_{2D}$, showing instead an increasing trend at low $n_{2D}$, reaching a maximum and then decreasing (see Fig.\ref{figure:IMT}e). This behavior can be observed in all ion-gated diamond surfaces \cite{YamaguchiJPSJ2013, TakahidePRB2014, AkhgarNanoLett2016, TakahidePRB2016, PiattiEPJST2019}. Since at low $T$ the electron-phonon scattering is suppressed, this behavior can be attributed to a crossover in the scattering with defects, specifically Coulomb scattering with the ions in the EDL. At low $n_{2D}$, the extra charge carriers introduced by the ionic gate improve the electrostatic screening, thus increasing the mobility \cite{YamaguchiJPSJ2013}. When $n_{2D}$ becomes too large, the additional charge carriers are no longer sufficient to effectively screen the extra disorder introduced by the ions and $\mu$ eventually decreases \cite{PiattiEPJST2019}. In the case of the (100) hydrogenated surface, the drop in $\mu$ is very sharp and might be assisted by an additional mechanism, namely the sudden increase in the interband scattering rate which occurs when high-energy sub-bands become filled at the increase of the doping level \cite{Gonnelli2dMater2017, PiattiApSuSc2017, PiattiNanoLett2018, YePNAS2011, GonnelliSciRep2015, PiattiJPCM2019}.

\section{Two-dimensional magnetotransport at low temperature}\label{sec:low_T}

Magnetotransport measurements are a powerful tool to investigate the physical properties of an electrically conducting system, since the dependence of $\sigma_{2D}$ on the applied magnetic field $B$ can provide crucial information concerning e.g. its dimensionality, the structure of its Fermi surface, the main sources of inelastic scattering and phase breaking,  and the interplay between its orbital and spin degrees of freedom. So far, there has not been a comprehensive examination of the magnetotransport properties of all gated diamond surfaces, with each work focusing on a specific feature of a specific surface. Nevertheless, the consensus appears to be that the magnetotransport in the (111) surface is dominated by weak localization (WL) onto which Shubnikov-de Haas (SdH) oscillations are superimposed in high-mobility samples \cite{TakahidePRB2014}; the magnetotransport in the (100) surface is instead dominated by weak anti-localization (WAL) caused by strong spin-orbit coupling (SOC), although the exact behavior and interpretation are debated between the different experimental reports \cite{AkhgarNanoLett2016, TakahidePRB2016, AkhgarPRB2019}. The magnetotransport properties of the (110) and polycrystalline surfaces have so far been unexplored. In this Section, we summarize the main results currently reported in the literature and suggest a possible explanation for the conflicting observations of Refs.\onlinecite{AkhgarNanoLett2016, TakahidePRB2016}.

\subsection*{(111) surface: Quantum oscillations}

The observation of quantum oscillations in B-doped diamond has been hampered by its low bulk mobility \mbox{($\mu\lesssim3\,\mathrm{cm^2V^{-1}s^{-1}}$) \cite{BorstPSSA1996, IshizakaPRL2007}}, mainly due to the very large B doping necessary to induce metallic behavior in diamond \cite{YokoyaNature2005, KleinPRB2007, KawanoPRB2010} as compared to B-doped silicon \cite{DaiPRB1992} or Ga-doped germanium \cite{ItohPRL1996}. Indeed, the first evidence for SdH oscillations in conducting diamond was reported in Ref.\onlinecite{TakahidePRB2014}, thanks to the large mobility (approaching $\mu\sim10^{2}\,\mathrm{cm^2V^{-1}s^{-1}}$, see Fig.\ref{figure:IMT}e) obtained in the ion-gated samples analyzed there. This is because the mobility in the gate-induced 2DHG can be significantly larger than the bulk value as long as the surface is atomically flat \cite{TakahidePRB2014}: This is necessary for the surface roughness to be smaller than the electrostatic screening length, which in gated diamond is expected to be of the order of $1-2\,\mathrm{nm}$ \cite{DankerlPRL2011, NebelDRM2004, EdmondsPRB2010, NakamuraPRB2013} for typical values of $n_{2D} \lesssim 10^{14}\,\mathrm{cm^{-2}}$ (larger $n_{2D}$ values can lead to strong deviations from the Thomas-Fermi approximation \cite{PiattiPRB2017, UmmarinoPRB2017, PiattiApSuSc2018nbn}).

\begin{figure}
\begin{center}
\includegraphics[keepaspectratio, width=1.0\columnwidth]{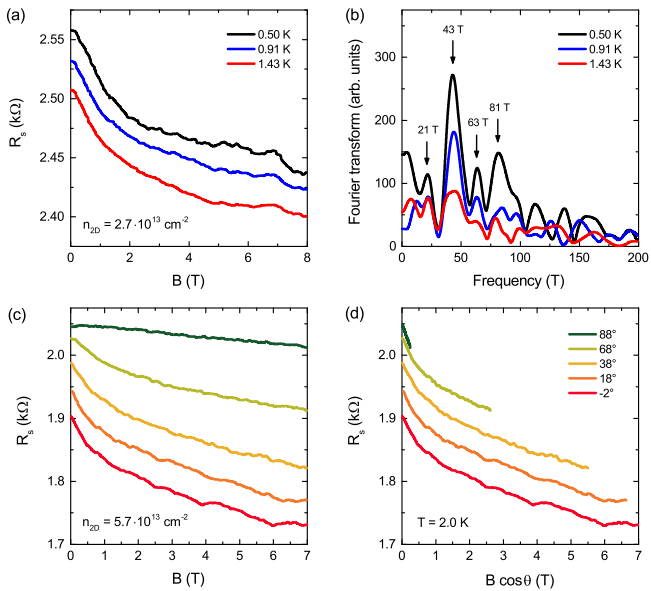}
\end{center}
\caption {
(a) Sheet resistance $R_s$ as a function of the magnetic field $B$ applied in the out-of-plane direction for different values of $T$ in a gated (111) diamond surface. Curves at $T = 0.91\,\mathrm{K}$ and $T = 1.43\,\mathrm{K}$ are shifted for clarity.
(b) Fourier spectra of the $B^{-1}$-dependence of the oscillations shown in (a). Frequencies at $21\,\mathrm{T}$, $43\,\mathrm{T}$, $63\,\mathrm{T}$ and $81\,\mathrm{T}$ are highlighted by arrows.
(c) $R_s$ vs. $B$ for another gated (111) diamond surface, for different angles $\theta$ between $B$ and the surface ($\theta = 0^\circ$ indicates that $B$ is perpendicular to the surface). Curves at $\theta = 18^\circ$, $\theta = 38^\circ$, $\theta = 68^\circ$ and $\theta = 88^\circ$ are shifted for clarity.
(d) Same data of (c) plotted vs. $B\mathrm{cos\theta}$, highlighting that the magnetotransport features depend only on the out-of-plane component of $B$.
All data are adapted from Ref.\onlinecite{TakahidePRB2014}.
} \label{figure:SdH}
\end{figure}

The low-$T$ sheet resistance $R_s$ of a high-mobility \mbox{($\mu = 91\,\mathrm{cm^2V^{-1}s^{-1}}$)} (111) gated surface is shown in Fig.\ref{figure:SdH}a as a function of the magnetic field intensity $B$, with $\vec{B}$ applied normal to the surface. $R_s$ decreases with increasing $B$ and shows a cusp-like behavior for  $B\lesssim 1\,\mathrm{T}$, suggesting WL behavior \cite{TakahidePRB2014}. The smaller quantum oscillations are observed on top of the WL background and are smoothed out with increasing $T$ \cite{TakahidePRB2014}. The Fourier spectra of the $B^{-1}$-dependence of the oscillations are shown in Fig.\ref{figure:SdH}b and exhibit peaks at $21\,\mathrm{T}$, $43\,\mathrm{T}$, $63\,\mathrm{T}$ and $81\,\mathrm{T}$. While the number of peaks and their positions were found to vary between devices and applied values of $V_G$, the sum of the corresponding sheet carrier densities was systematically smaller than the value of $n_{2D}$ determined from Hall effect. Ref.\onlinecite{TakahidePRB2014} attributed this feature either to the undetected presence of Fermi surfaces with large scattering rates and/or effective masses, or -- more probably -- to the well-known inhomogeneous carrier density typical of ion-gated surfaces \cite{JoNanoLett2015, RenNanoLett2015, DeziPRB2018}: the SdH signal would arise only from the small, high-$\mu$ regions embedded in a low-$\mu$ background, consistent with their small intensity.

{\color{blue}The values of the hole effective masses, determined from the $T$-dependence of the SdH frequencies, could be grouped in two separate ranges ($m^*/m_0 = 0.17-0.36$ and $m^*/m_0 = 0.57-0.78$, $m_0$ being the electron mass) \cite{TakahidePRB2014}, which} were found to be compatible with those of the light-hole and heavy-hole bands of (111) pristine single-crystal diamond \cite{NakaPRB2013}, with multiple frequencies often exhibiting the same effective mass. {\color{blue}Surprisingly, only the first of these two ranges is compatible with the effective masses calculated ab initio for the gated (111) surface \cite{RomaninApSuSc2019}.} 

Moreover, the quantum scattering rate was found to be systematically lower (by one order of magnitude) than the transport scattering rate. This feature, together with the presence of multiple SdH frequencies, is again consistent with spatial inhomogeneities in the gate-induced hole density, {\color{blue}since it points to the existence of spatially distinct regions with different local values of $n_{2D}$ and with} a local $\mu$ exceeding $10^{3}\,\mathrm{cm^2V^{-1}s^{-1}}$ at low $T$ \cite{TakahidePRB2014}.

Finally, the magnetoresistance of the gated (111) surface was found to be strongly dependent on the angle $\mathrm{\theta}$ between $B$ and the out-of-plane direction. As $B$ was tilted towards the in-plane direction ($\mathrm{\theta}\rightarrow90^{\circ}$, see Fig.\ref{figure:SdH}c), the negative magnetoresistance nearly disappeared and the nodes of the SdH oscillation shifted towards larger values of $B$. Indeed, Ref.\onlinecite{TakahidePRB2014} showed that the magnetoresistance signal depended only on the $B$ component perpendicular to the surface ($B\mathrm{cos\theta}$, see Fig.\ref{figure:SdH}d), providing a direct evidence for the 2D character of the Fermi surface of the gate-induced 2DHGs.

\subsection*{(100) surface: Spin-orbit coupling}

\begin{figure*}
\begin{center}
\includegraphics[keepaspectratio, width=1.0\textwidth]{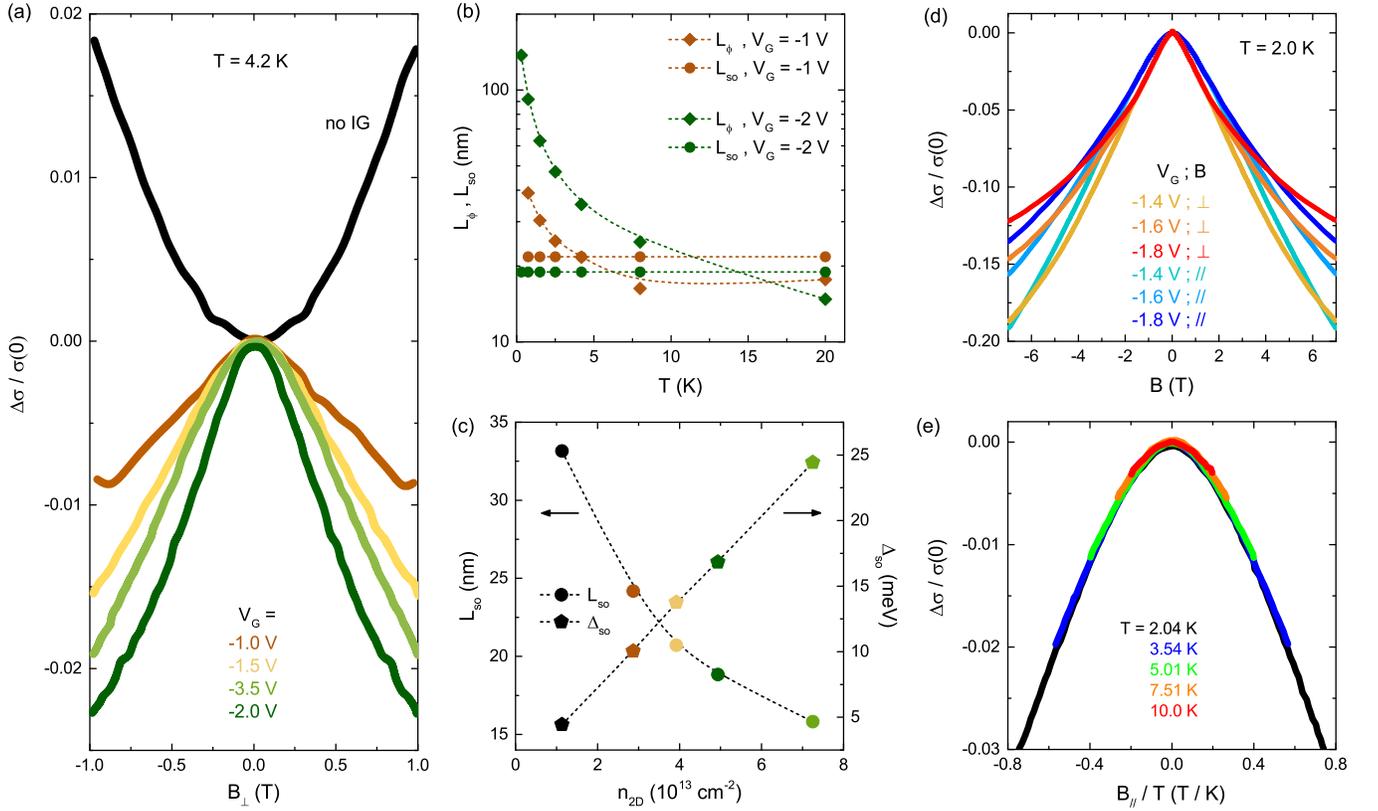}
\end{center}
\caption {
(a) In-plane magnetoconductance ratio \mbox{$[\sigma (B) - \sigma(0)] / \sigma(0)$} vs. out-of-plane magnetic field $B_\perp$, for different values of $V_G$ applied at a (100) diamond surface. The label \mbox{``no IG"} marks the curve in absence of the ionic gate.
(b) $T$-dependences of the phase-coherence length $L_\phi$ (diamonds) and spin-coherence length $L_{so}$ (circles) for two different values of $V_G$. 
(c) Spin-coherence length $L_{so}$ (circles, left scale) and spin-orbit splitting $\Delta_{so}$ (pentagons, right scale) as a function of $n_{2D}$. Dashed lines in both (b) and (c) are guides to the eye.
(d) In-plane magnetoconductance ratio \mbox{$[\sigma (B) - \sigma(0)] / \sigma(0)$} as a function of either the out-of-plane magnetic field $B_\perp$ (orange-red lines) or the in-plane magnetic field $B_{\parallelslant}$ (green-blue lines) for different values of $V_G$.
(e) In-plane magnetoconductance ratio vs. $B_{\parallelslant}/T$. Data for different values of $V_G = -1.4\,\mathrm{V}\,,-1.6\,\mathrm{V}\,,-1.8\,\mathrm{V}$ and $T$ (as indicated in the legend) all collapse on the same curve.
Data in (a), (b) and (c) are adapted from Ref.\onlinecite{AkhgarNanoLett2016}. Data in (d) and (e) are adapted from Ref.\onlinecite{TakahidePRB2016}.
} \label{figure:SOC}
\end{figure*}

The magnetotransport properties of the gated (100) diamond surface were first investigated in Refs.\onlinecite{AkhgarNanoLett2016, TakahidePRB2016}. The two works focused on close but non-overlapping ranges of doping: Ref.\onlinecite{AkhgarNanoLett2016} investigated magnetotransport in the ungated surface, $n_{2D}\simeq 1.1\cdot10^{13}\,\mathrm{cm^{-2}}$, and in the gated surface in the range $n_{2D}\simeq 2.4-7.2\cdot10^{13}\,\mathrm{cm^{-2}}$. Ref.\onlinecite{TakahidePRB2016} focused instead on the gated surface only and in the narrow range $n_{2D}\simeq 1.15-1.72\cdot10^{13}\,\mathrm{cm^{-2}}$. Surprisingly, despite the same device platform and the moderate difference between the doping values, the results (and their interpretation) of the two works were at variance with one another, only agreeing on the fundamental role played by SOC in the observed behaviors. Moreover, the negative magnetoconductance observed in Refs.\onlinecite{AkhgarNanoLett2016, TakahidePRB2016} in the (100) diamond surface is in stark contrast with the positive magnetoconductance observed in Ref.\onlinecite{TakahidePRB2014} in the (111) surface, and provides evidence of a wide difference in the role played by SOC between the two surfaces.

Ref.\onlinecite{AkhgarNanoLett2016} reported that, for $T\lesssim20\,\mathrm{K}$, the $T$-dependent electric transport was dominated by hole-hole interactions (HHI), as evidenced by the logarithmic dependence of the transverse resistance on $T$. Indeed, when the small HHI contribution \cite{GohPRB2008} was subtracted from $R_s$, the resulting corrected longitudinal sheet resistance was found to be nearly $T$-independent, both in the presence and in the absence of the ionic gate. The authors then conducted the subsequent analysis on the HHI-corrected magnetotransport data, evidencing a $T$- and $n_{2D}$-dependent crossover from WL to WAL caused by a gate-tuned SOC strength \cite{AkhgarNanoLett2016}. The $n_{2D}$-dependent crossover is exemplified in Fig.\ref{figure:SOC}a at a fixed $T = 4.2\,\mathrm{K}$. In the absence of the ionic gate, the positive magnetoconductivity ratio $\Delta \sigma / \sigma (0) = [\sigma(B) - \sigma(0)] / \sigma(0)$ is due to WL. When $n_{2D}$ is increased, $\Delta \sigma / \sigma (0)$ becomes increasingly negative, signaling the presence of WAL. A similar crossover is induced by decreasing $T$, as already reported for the ungated surface \cite{EdmondsNanoLett2015}.

Ref.\onlinecite{AkhgarNanoLett2016} attributed the observed WAL to a gate-tunable SOC of dominant Rashba type -- induced by the broken inversion symmetry in the 2DHG caused by the strong interface electric field -- and fitted the magnetoconductivity data with the corresponding theoretical expression (Rashba splitting proportional to the wave vector to the third power) \cite{IordanskiiJAP1994, KnapPRB1996} to extract the phase and spin-orbit coherence lengths $L_\phi$ and $L_{so}$ as a function of $n_{2D}$ and $T$. In the presence of a finite SOC, the spin of the charge carriers precedes as they are scattered in closed loops, the sign of this precession being opposite for the two loop directions: This changes the constructive interference of coherent carrier backscattering into destructive interference \cite{HikamiPTP1980}. Thus, WAL is observed whenever $L_\phi$ exceeds $L_{so}$ and the gate-dependent crossover from WL to WAL was ascribed to the increase of $n_{2D}$ which extends the $T$ range where $L_\phi \geq L_{so}$ (see Fig.\ref{figure:SOC}b). Instead, the $T$-dependent crossover (not shown here) was attributed to Nyquist hole-hole scattering suppressing $L_\phi$ as $T^{-1/2}$, eventually pushing it below the $T$-independent $L_{so}$. From the $n_{2D}$-dependence of $L_{so}$, the authors also extracted the dependence of the Rashba spin-orbit splitting $\Delta_{so}$, which was found to increase by over 4 times as $n_{2D}$ was increased (see Fig.\ref{figure:SOC}c) even though the intrinsic Rashba coupling strength decreased (anomalous / negative differential Rashba splitting) \cite{WinklerPRB2002, HabibAPL2004}. Notably, the maximum gate-induced $\Delta_{so}\simeq 24.5\,\mathrm{meV}$ is larger than those reported for most 2D hole systems \cite{AkhgarNanoLett2016}, and is comparable to those of 2D electron systems such as HgTe quantum wells \cite{GuiPRB2004} or the high-energy valleys of electron-doped MoS\ped{2} \cite{PiattiJPCM2019}.

Ref.\onlinecite{TakahidePRB2016} also reported WAL behavior in gated (100) surfaces at $T\lesssim 10\,\mathrm{K}$. As in Ref.\onlinecite{AkhgarNanoLett2016}, the negative magnetoconductance was orders of magnitude larger than the classical orbital contribution \cite{KittelBook}, suggesting the dominant role of the spin degree of freedom. This was confirmed by $\Delta \sigma / \sigma (0)$ being measured not only as a function of the out-of-plane magnetic field $B_\perp$ (as in Ref.\onlinecite{AkhgarNanoLett2016}), but also of the in-plane magnetic field $B_{\parallelslant}$, and its magnitude found to be weakly dependent on the direction of $B$ (see Fig.\ref{figure:SOC}d) \cite{TakahidePRB2016}. However, no crossover between WL and WAL was reported, and $\Delta \sigma / \sigma (0)$ was nearly insensitive to changes in $n_{2D}$ at low $B$, in sharp contrast to the behavior reported in Ref.\onlinecite{AkhgarNanoLett2016} and shown in Fig.\ref{figure:SOC}a. Most notably, all the in-plane magnetoconductance data reported in Ref.\onlinecite{TakahidePRB2016} collapse on the same universal curve when plotted as a function of $B_{\parallelslant} / T$ (see Fig.\ref{figure:SOC}e), and the out-of-plane magnetoconductance data do the same as a function of $B_{\perp} / T^{1.32}$ (not shown), irrespectively of the values of $n_{2D}$ and $T$ \cite{TakahidePRB2016}.

Several theories have been developed for 2D conducting systems in the presence of SOC and Zeeman splitting \cite{HikamiPTP1980, KawabataJPSJ1981, MaekawaJPSJ1981, LeePRB1982, CastellaniPRB1998} which predict \mbox{$\Delta \sigma \propto - (B/T)^2$}, but they all have strong difficulties in reproducing the observed \mbox{$\Delta \sigma / \sigma \propto - (B/T)^2$} \cite{TakahidePRB2016}. This second dependence has been instead predicted in the hopping regime in the presence of localized spins \cite{AgrinskayaSSC1998}, where the Pauli exclusion principle suppresses the carrier hopping probability to those sites where the localized spin is aligned to that of the charge carrier. Although the gated (100) diamond surface is far away from the hopping regime (as discussed in Ref.\onlinecite{TakahidePRB2016} and Section \ref{figure:IMT}), and in the absence of a fully satisfactory theory, the authors proposed that a similar process may be at play in their devices due to the remaining dangling bonds after the (100) surface reconstruction, which provide localized spins and constitute a known source of magnetic noise \cite{GrinoldsNatNano2014, RosskopfPRL2014, MyersPRL2014}.

Several possible reasons can be cited for the discrepancy between the results of Refs.\onlinecite{AkhgarNanoLett2016, TakahidePRB2016}. First, the data in Ref.\onlinecite{TakahidePRB2016} were not corrected for HHI, as were those of Ref.\onlinecite{AkhgarNanoLett2016} instead. On the other hand, Ref.\onlinecite{AkhgarNanoLett2016} did not perform measurements for in-plane magnetic fields, and Ref.\onlinecite{TakahidePRB2016} pointed out how WAL theories cannot account for non-saturating magnetoconductance even at $2\,\mathrm{T}$, which is observed in the data of Ref.\onlinecite{AkhgarNanoLett2016} at low $n_{2D}$. A strong difference in the density of dangling bonds in the crystals employed by the two groups cannot be ruled out but seems unlikely, given that they act as trap states and the mobility of the two sets of samples is comparable (see Fig.\ref{figure:IMT}e). Overall, we deem the most likely explanation to lie in the different doping regimes investigated by the two works: Since Ref.\onlinecite{AkhgarNanoLett2016} focuses on doping levels at least twice as large as those of Ref.\onlinecite{TakahidePRB2016}, the dangling-bond induced trap states would become more filled (reducing the density of unpaired localized spins), while the Rashba splitting increases due to the increase of the interface electric field. The WAL signal would then crossover from being dominated by spin-spin interactions at low doping (responsible for the universal \mbox{$\Delta \sigma / \sigma \propto - (B/T)^2$} behavior observed in Ref.\cite{TakahidePRB2016}), to Rashba SOC at high doping (responsible for the more standard, Hikami-like magnetoconductance observed in Ref.\onlinecite{AkhgarNanoLett2016}).

\subsection*{(100) surface: $g$-factor and well-width fluctuations}

Further insight into the physical properties of the gate-induced 2DHG at the (100) surface can be gained by investigating its magnetoconductance upon the simultaneous application of finite $B_\perp$ and $B_{\parallelslant}$. Namely, one can obtain estimations for the carrier $g$-factor, which measures the strength of the coupling between the carrier spin and a magnetic field, and the mean square well-width roughness $d^2$, which measures how much the local depth of the 2DHG from the surface deviates from its average value (and, thus, from an ideal 2D system). This was first done in Ref.\onlinecite{AkhgarPRB2019}, following the methods developed in Refs.\onlinecite{MinkovPRB2004, AkhgarAPL2018}, for $n_{2D} \lesssim 4.35\cdot 10^{13}\,\mathrm{cm^{-2}}$.

\begin{figure}
\begin{center}
\includegraphics[keepaspectratio, width=1.0\columnwidth]{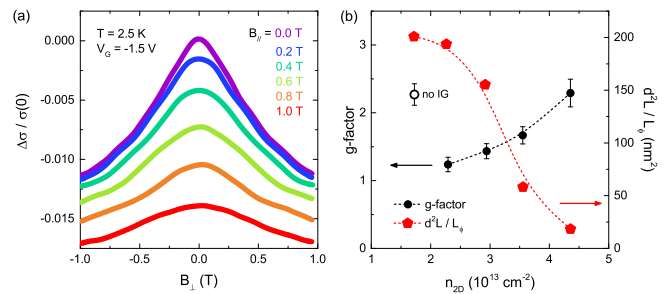}
\end{center}
\caption {
(a) In-plane magnetoconductance ratio \mbox{$[\sigma (B_\perp , B_{\parallelslant}) - \sigma(0,0)] / \sigma(0,0)$} vs. out-of-plane magnetic field $B_\perp$, for different values of the in-plane magnetic field $B_{\parallelslant}$, in a gated (100) diamond surface.
(b) In-plane $g$-factor (black circles, left scale) and roughness parameter $d^2L$ divided by the phase-breaking length $L_\phi\sim30\,\mathrm{nm}$ (red pentagons, right scale) as a function of $n_{2D}$. The label \mbox{``no IG"} marks the value in absence of the ionic gate. Dashed lines are guides to the eye.
All data are adapted from Ref.\onlinecite{AkhgarPRB2019}.
} \label{figure:g_and_fluct}
\end{figure}

As shown in Fig.\ref{figure:g_and_fluct}, a quenching of the WAL signal vs. $B_\perp$ with increasing $B_{\parallelslant}$ was observed, which can be accounted for by introducing two additional phase-breaking fields, $\Delta_r$ and $\Delta_s$, in the standard Hikami formula for WAL in 2D \cite{AkhgarPRB2019, MinkovPRB2004}. $\Delta_r$ is proportional to $\frac{d^2L}{l}B_{\parallelslant}^2$, where $L$ is the correlation length of the well-width fluctuations and $l$ is the mean free path, and represents the fact that, when $d^2$ is finite, charge carriers scattering in closed loops are no longer strictly confined to a 2D sheet and thus acquire an additional Aharonov-Bohm phase. $\Delta_s$ is proportional to $g\mu_BB_{\parallelslant}^2$ ($\mu_B$ being the Bohr magneton) and is due to the Zeeman effect. Thus, Ref.\onlinecite{AkhgarPRB2019} determined the $g$-factor and $d^2L$ (see Fig.\ref{figure:g_and_fluct}b) by linearly fitting the $B_{\parallelslant}^2$-dependence of $\Delta_s$ and $\Delta_r$ as obtained from fitting the Hikami model to the experimental data.

The $g$-factor was found to sharply decrease upon the deposition of the ionic liquid, and then to increase monotonically up to $2.3$ with increasing $n_{2D}$. Since deviations of $g$ from the free-electron value ($g=2$) arise from changes in the bandstructure, in the absence of an accurate model the authors proposed that the modulation of $g$ might be due to a combination of valence band filling and hybridization of the light-hole and heavy-hole sub-bands \cite{AkhgarPRB2019}. The roughness parameter $d^2L/L\ped{\phi}$, on the other hand, was found to decrease monotonically with increasing $n_{2D}$, eventually approaching the mean-square surface roughness $\simeq 1.2\,\mathrm{nm^2}$ at large $n_{2D}$. This implies that, in the measured range of $n_{2D}$, the gate-induced 2DHG becomes more homogeneous as more holes become trapped in the potential well \cite{AkhgarPRB2019} and is consistent with the known behavior of ion-gated surfaces \cite{RenNanoLett2015}.

\section{Superconductivity}\label{sec:supercond}

Superconductivity can be induced in bulk diamond when the B-dopant concentration becomes large enough, with typical values for the SC critical temperature $T_c \simeq 4\,\mathrm{K}$ for hole densities $n_{3D}\approx 10^{22}\,\mathrm{cm^{-3}}$ \cite{EkimovNature2004,BustarretPRL2004,BustarretPSSA2008}. B-doped diamond films grown homoepitaxially on intrinsic single crystals feature maximum $T_c\simeq 3\,\mathrm{K}$ for the (100) orientation and $T_c\simeq 7\,\mathrm{K}$ for the (111) orientation \cite{TakanoJPCM2009}. In the latter case, a further improvement to $T_c\simeq 10\,\mathrm{K}$ has been obtained in highly-ordered films \cite{OkazakiAPL2015}. Indeed, theoretical predictions indicate \cite{BlasePRL2004, BoeriPRL2004, LeePRL2004, BoeriJPCS2006, GiustinoPRL2007, BlaseNatMater2009} that further incrementing the B content in the compact diamond crystal structure should allow obtaining $T_c\sim 40\,\mathrm{K}$. However, the successful achievement of such a large $T_c$ is hampered by the limited solubility of B in the diamond lattice, as well as by the disorder introduced by the substitution process \cite{BustarretPSSA2008, TakanoAPL2004, OkazakiAPL2015}. Surprisingly, little theoretical effort has been devoted to explore the possibility of superconductivity induced by field-effect doping in diamond, which should be much less prone to the introduction of extrinsic disorder. 

\subsection*{(110) surface}

The influence of field-effect doping on the hydrogenated (110) diamond surface has been investigated theoretically in Refs.\onlinecite{NakamuraPRB2013, SanoPRB2017}. In these works, the metallic gate was modeled by a homogeneous planar distribution of charges. The resulting uniform electric field is normal to the diamond surface and becomes screened inside the material by a self-consistent redistribution of the charge density. In principle, this allows to induce both electron and hole surface doping by selecting the sign of the applied field (conventionally, the reference system is chosen so that a negative sign of the applied field corresponds to hole doping).

\begin{figure}
\begin{center}
\includegraphics[keepaspectratio, width=1.0\columnwidth]{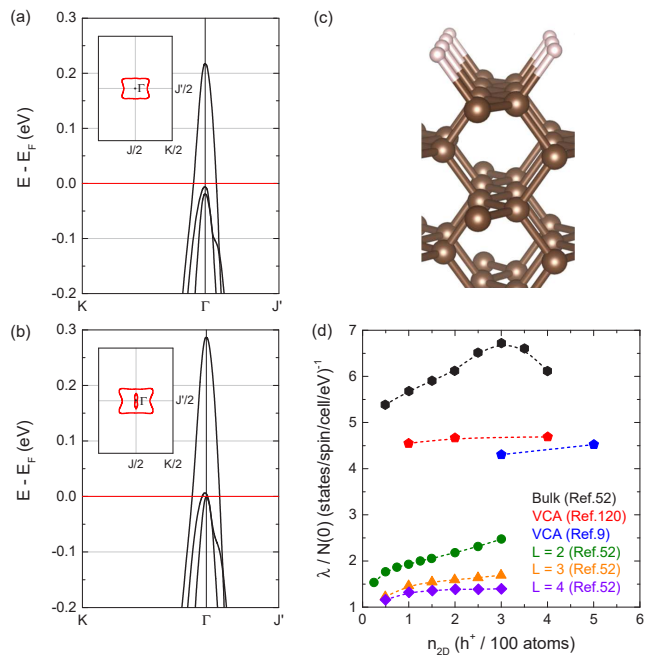}
\end{center}
\caption {
(a) Electronic bandstructure of the ion-gated, hydrogen-terminated (110) $1\times1$  diamond surface for $n_{2D}=2.84\cdot10^{13}\,\mathrm{cm^{-2}}$. The inset shows the first Brillouin zone and the Fermi surface.
(b) Same as in (a) but for $n_{2D}=5.68\cdot10^{13}\,\mathrm{cm^{-2}}$.
(c) Ball-and-stick model of the crystal lattice. Large brown spheres and small pink spheres are carbon and hydrogen atoms, respectively.
(d) Electron-phonon coupling constant normalized to the DOS at the Fermi level, $\lambda/N(0)$, as a function of $n_{2D}$ in units of induced carriers per 100 atoms.
Data in (a,b) are adapted from Ref.\onlinecite{NakamuraPRB2013}; note that only the uppermost valence bands are shown.  Data in (d) are adapted from Refs.\onlinecite{SanoPRB2017, BoeriPRL2004, MaPRB2005}.
} \label{figure:SC_110}
\end{figure}

Ref.\onlinecite{NakamuraPRB2013} first computed the electronic structure of the (110) surface -- modeled with a slab of 13 carbon layers terminated by hydrogen on both sides -- in the absence of the electric field. The DFT calculations were performed with the full-potential linearized augmented plane-wave method \cite{WimmerPRB1981, WeinertPRB1982, NakamuraPRL2009, WeinertPCM2009}. The as-cleaved surface has only one dangling bond per surface carbon atom. Thus, as in the case of the (111) surface, hydrogen can saturate all the missing bonds, stabilizing the $1\times1$ cell and avoiding any reconstruction. The hydrogenated (110) diamond surface is shown in Fig.\ref{figure:SC_110}c, and the corresponding point group is $C_{2v}$ with a two-fold rotation axis in [110]. In the absence of external doping, the surface is insulating with a direct band gap of $\approx 2.5\,\mathrm{eV}$ at the $\Gamma$ point of the first Brillouin zone. The bonding and antibonding orbitals of hydrogen are several electronvolts below and above the valence-band maximum, respectively.

When a finite negative electric field was applied to the surface, hole carriers were induced in the first few carbon layers ($\sim 5-10\,\mathrm{\AA}$ below the surface): The Fermi level crossed the first valence bands and the surface became metallic. The gate electric field had no significant effect on the crystal lattice, only slightly reducing the bonding length between the surface atoms. The authors of Ref.\onlinecite{NakamuraPRB2013} focused on two values of the electric field ($E = -0.5\,\mathrm{V/\AA}$ and $E = -1.0\,\mathrm{V/\AA}$), corresponding to the two hole doping levels $n_{2D} = 2.84\cdot10^{13}\,\mathrm{cm^{-2}}$ and $n_{2D} = 5.68\cdot10^{13}\,\mathrm{cm^{-2}}$. At $n_{2D} = 2.84\cdot10^{13}\,\mathrm{cm^{-2}}$, the Fermi level crossed only the uppermost valence band (see Fig.\ref{figure:SC_110}a), leading to a single holonic pocket at the zone center. On increasing the doping to $n_{2D} = 5.68\cdot10^{13}\,\mathrm{cm^{-2}}$ the Fermi level shifted to lower energies. This resulted in the increase of the Fermi surface size and in the crossing of the second valence band as well (see Fig.\ref{figure:SC_110}b), leading to the emergence of a second holonic pocket at zone center. Notably, the filled valence bands are also \textit{surface-bound hole states} in the sense that the charge distributions of their wavefunctions are spatially distributed in planar sheets \cite{SanoPRB2017}, and the confinement of the induced holonic carriers within few atomic layers from the surface results in a value of $n_{3D}$ exceeding the one responsible for the onset of SC in B-doped bulk diamond \cite{NakamuraPRB2013}. The authors then computed the electron-phonon coupling $\lambda$ in the rigid-muffin-tin approximation \cite{GaspariPRL1972, SkriverPRB1990, RhimPRB2007}, obtaining $\lambda=0.18$ at $n_{\text{2D}}=2.84\cdot10^{13}\,\mathrm{cm^{-2}}$ and $\lambda=0.47$ at $n_{\text{2D}}=5.68\cdot10^{13}\,\mathrm{cm^{-2}}$. Even though they did not attempt to compute the $T_c$, the fact that bulk diamond features a comparable $\lambda=0.51$ for a $12\%$ B concentration \cite{NakamuraPRB2013} led them to suggest that the value of $\lambda$ at $n_{\text{2D}}=5.68\cdot10^{13}\,\mathrm{cm^{-2}}$ should be sufficient to trigger the onset of SC in the surface accumulation layer.

The field-induced SC transition temperature was calculated in Ref.\onlinecite{SanoPRB2017} by DFT, with the pseudopotential plane-wave method \cite{GiannozziPCM2009}, and in the same geometry as in Ref.\onlinecite{NakamuraPRB2013}. Fig.\ref{figure:SC_110}d summarizes the calculated dependence of $\lambda/N(0)$ (that is the electron-phonon coupling constant normalized to the DOS at the Fermi level, and corresponds to the attractive potential in BCS theory, $V$) on increasing $n_{2D}$. The authors computed $\lambda/N(0)$ for B-doped bulk diamond (black hexagons) and for three ion-gated hydrogenated (110) slabs composed of two ($L = 2$, green circles), three ($L = 3$, orange triangles) and four ($L = 4$, violet diamonds) carbon layers as a function of the surface charge density. The results of DFT in virtual-crystal approximation for bulk B-doped diamond from Refs. \onlinecite{BoeriPRL2004} (blue pentagons) and \onlinecite{MaPRB2005} (red pentagons) are reported in Fig.\ref{figure:SC_110}d for comparison. By extrapolating the (almost doping-independent) resulting values of $\lambda/N(0)$ to a slab made up of a larger number of layers, the authors estimate the value of $V$ in the slab to be $V \simeq 1.1\,\mathrm{(states/spin/cell/eV)^{-1}}$. From here, and using a step-like approximation of the DOS, they estimated  $\lambda$ to increase from 0.08 (when only the first hole pocket is present) to 0.3 (when the second hole pocket forms at the zone center). The $T_{\mathrm{c}}$ of the the gated (110) surface was then estimated by using the  semi-empirical McMillan equation \cite{McMillanPR1968, AllenPRB1975}:

\begin{equation}
\label{eq:AllenDynes}
T_{\mathrm{c}}=\frac{\omega_{\text{log}}}{1.2}\exp{\Bigl\{-\frac{1.04(1+\lambda)}{\lambda-\mu^*(1+0.62\lambda)}\Bigr\}}
\end{equation}
where $\omega_{\text{log}}\sim1200$ K is the logarithmic averaged phonon frequency and $\mu^*\sim0.1$ is the typical Morel-Anderson pseudopotential of B-doped bulk diamond \cite{BoeriPRL2004, KleinPRB2007, BustarretPC2015, MaPRB2005}. 
The result of the whole procedure indicates the existence of critical value of the induced hole density $n_{2D,c}\approx2.3\cdot10^{13}\,\mathrm{cm^{-2}}$ above which a finite $T_{\mathrm{c}} \simeq 1$ K appears. For $n_{2D}<n_{2D,c}$, $\lambda\approx0.1$ and $T_c\rightarrow0$. The onset of superconductivity thus corresponds to a sudden jump in the DOS, $N(0)$, that occurs when the second valence band crosses the Fermi level and that strongly boosts the electron-phonon coupling to $\lambda\approx0.3$.  The situation is similar to that observed, for instance, in MoS$_2$ \cite{PiattiNanoLett2018,PiattiJPCM2019} where the superconducting transition does not appear until the Fermi level crosses both the spin-orbit split sub-bands at the $Q$ point.

{\color{blue}\subsection*{(111) surface}

The possible occurrence of superconductivity due to field-effect doping was also investigated in Ref.\onlinecite{RomaninApSuSc2019} on the hydrogenated (111) diamond surface (see Fig.\ref{figure:bands}c). In this work, ab-initio plane-wave pseudopotential density functional theory computations were carried out in the proper field-effect geometry \cite{SohierPRB2017} (i.e. by taking into account the screening of the electric field in a self-consistent way on structure relaxation, electronic and vibrational properties and electron-phonon coupling). Three different hole doping values were investigated:  $n_{\text{2D}}=2.84\cdot10^{13}\,\mathrm{cm}^{-2}$, $n_{\text{2D}}=1.96\cdot10^{14}\,\mathrm{cm}^{-2}$ and $n_{\text{2D}}=6.00\cdot10^{14}\,\mathrm{cm}^{-2}$.

By letting the structure relax in the presence of the electric field, the authors showed that atomic positions were not dramatically affected and the only effect was just a slight variations of bond lengths. The electronic structure however was shown to be strongly dependent on the field-effect doping: as a matter of fact, on increasing doping the Fermi level is crossed by bands that change their character from a bulk-like to a surface-like one. Moreover, the presence of the electric field in the direction perpendicular to the surface breaks inversion symmetry along the axis perpendicular to the surface and this results in a lifting of degeneracies among bands (see Fig.\ref{figure:bands}d and Fig.\ref{figure:SC_111}a  for comparison).

Electron-phonon interactions as a function of doping were initially computed only at the center of the Brillouin zone (i.e. $\bf{q}=\bf{\Gamma}$). The resulting electron-phonon coupling constants $\lambda$ and logarithmic averaged frequencies $\omega_{\text{log}}$ are reported in Fig.\ref{figure:SC_111}b. The transition temperature was then computed using the semi-empirical McMillan equation (Eq.\ref{eq:AllenDynes}), showing that at high hole doping (i.e. $n_{\text{2D}}=6.00\cdot10^{14}$ cm$^{-2}$) superconductivity appears with $T_{c}\in[57.20\,\mathrm{K},63.14\,\mathrm{K}]$ ($\mu^*\in[0.13,0.14]$), and that it is mainly due to planar vibrations and -- to a lesser extent -- to out-of-plane vibrations. For this case the authors found $\lambda=1.09$ and $\omega_{\rm log}=629.94$ cm$^{-1}$.

However, a more accurate analysis of electron-phonon interactions, using a Wannier interpolation scheme \cite{CalandraPRB2010}, showed that the electron-phonon coupling constant is renormalized by $\approx30\%$ giving $\lambda=0.81$, while the logarithmic averaged frequency is $\omega_{\rm log}\approx 670$ cm$^{-1}$. Then, the superconductive critical temperature was re-evaluated both using Eq.\ref{eq:AllenDynes} and, more accurately, by solving the isotropic linearized single-band Eliashberg equations \cite{Eliashberg1,Eliashberg2,AllenMitrovic}. In the first case the transition temperature turned out to be $T_{c}\in[29\,\mathrm{K},35\,\mathrm{K}]$ (with $\mu^*\in[0.13,0.14]$), while the resulting critical temperature obtained by solving the Eliashberg equations is $T_{c}\approx36$ K (with $\mu^*=0.17$, which is the value necessary to reproduce the experimental $T_c$ of B-doped diamond within the Eliashberg approach).

\begin{figure}
\begin{center}
\includegraphics[keepaspectratio, width=1.0\columnwidth]{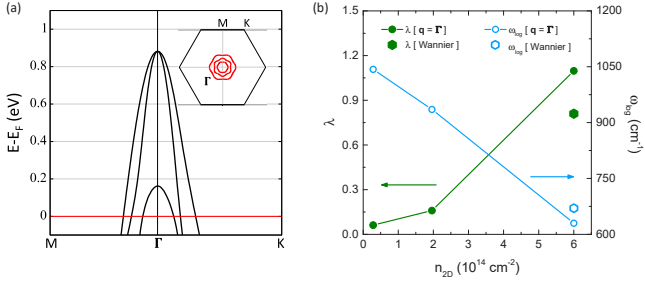}
\end{center}
\caption {{\color{blue}
(a) Electronic bandstructure of the ion-gated, hydrogen-terminated (111) $1\times1$  diamond surface for $n_{2D}=6.00\cdot10^{14}\,\mathrm{cm^{-2}}$. The inset shows the first Brillouin zone and the Fermi surface.
(b) Electron-phonon coupling constants $\lambda$ (filled symbols, left axis) and averaged logarithmic frequencies $\omega_{\rm log}$ (hollow symbols, right axis) as a function of hole doping $n_{2D}$ computed at $\bf{q}=\bf{\Gamma}$ (filled and hollow circles) and through a Wannier interpolation scheme (filled and hollow hexagons).
Data in (a,b) are adapted from Ref.\onlinecite{RomaninApSuSc2019}.}
} \label{figure:SC_111}
\end{figure}

As a final remark, in Ref.\onlinecite{RomaninApSuSc2019} the authors explain that the electron-phonon coupling arises partly from interband scattering, making the system multiband in nature. Therefore, the values of $T_{c}$ reported in their work might be an underestimation of the real $T_{c}$.}

\section{Summary and outlook}\label{sec:summary}

{\color{blue}In this short review,} we have discussed the electronic structure and the electric transport properties of the two-dimensional hole gas induced at the surface of ion-gated diamond. The role of the different surface orientations in determining the valence-band structure of hydrogen-terminated diamond has been examined based on the results of DFT calculations. We have discussed how the ionic gating technique can be used to control the hole density in the accumulation layer at the (111), (100) and nanocrystalline surfaces, and whether and how hole transport can be tuned across the insulator-to-metal transition, highlighting the different roles of intrinsic and gate-induced disorder. We have further reviewed the low-temperature magnetotransport properties of the (111) and (100) ion-gated  surfaces, discussing two-dimensional weak localization and Shubnikov-de Haas oscillations in the (111) surface and different regimes of weak anti-localization and gate-tunable spin-dependent transport in the (100) surface. Finally, we have discussed the electronic structure and superconductivity predicted for the hydrogen-terminated (110) {\color{blue}and (111) surfaces} on the basis of DFT calculations.

At the end we can suggest some promising avenues for future investigations. From the theoretical point of view, the spin-dependent transport in the hydrogenated (100) surface still lacks a comprehensive and quantitative explanation for the observed features. Additionally, the possibility and requirements for gate-induced superconductivity have not yet been investigated in the (100) surface. Conversely, from the experimental point of view, the electric transport properties of the ion-gated (110) surface are still to be explored, a promising task owing to the predicted superconductivity. Finally, superconducting behavior has not been observed in either the (111) or (100) gated surfaces, most likely due to the insufficient hole density achieved so far. Since larger gate-induced hole densities have been reported in nanocrystalline surfaces upon B-doping, we deem the combination of B-doping and ionic gating in the same epitaxial film to be a promising avenue to finally reach a gate-tunable superconducting state at the surface of diamond.

\section*{Acknowledgments}
We thank M. Calandra and A. Pasquarelli for fruitful scientific discussions. Atomic structures were rendered with VESTA \cite{VESTA}.

\end{document}